# Stroke Fragmentation based on Geometry Features and Hidden Markov Model

Guihuan Feng, Christian Viard-Gaudin, Technical Report, IRCCyN Nantes/IVC

**ABSTRACT**
Stroke fragmentation is one of the key steps in pen-based interaction. In this letter, we present a unified HMM-based stroke fragmentation technique that can do segment point location and primitive type determination simultaneously. The geometry features included are used to evaluate local features, and the HMM model is utilized to measure the global drawing context. Experiments prove that the model can efficiently represent smooth curves as well as strokes made up of arbitrary lines and circular arcs.

**Keywords**
Pen-based interaction, stroke fragmentation, hidden markov model, hand-drawn sketch.

## 1. INTRODUCTION

Stroke fragmentation is a perceptual analysis of ink. It tries to cluster stroke points into geometrically salient primitives, such as line segments and circular arcs [4]. There is no doubt that it is one of the key steps in pen-based interaction. First, computation based on raw points will be tedious and time consuming; second, comparatively, primitives contain more meaningful geometry information than points do; at last, experiments proved that robust stroke fragmentation shall improve the performance of sketch understanding systems [5]. Whereas, due to the informality of sketches and users' inadaptability to the input devices, stroke fragmentation still remains to be a challenge.

Segment point location and primitive type determination are the two main issues in stroke fragmentation. Segment point location aims at figuring out how many segments are there and where are their starting and ending points; and primitive type determination is used to label each of the segments. Most existing works [1,5,8,11] are based on curvatures and speeds, points whose those values are below a given threshold are taken as segment points. The differences lie only on the selection of the filtering techniques. Such methods are easy to be implemented and they perform well in many applications. But as they focus only on local contexts, sometimes may be susceptible to over- and under- segmentation of strokes. Furthermore, it cannot deal with smooth curves, such as "J" and "S", correctly. Although Hse combines the two issues together by Dynamic programming (DP) algorithm [4], as the approach is based on symbol recognition, it's difficult to extend in both type of the stroke and type of symbol.

Hidden Markov Model was first introduced in 1960s. As it is especially good at modeling sequential and temporal phenomena, it has received many achievements in applications, such as speech recognition [7], gesture recognition [6] and handwritten document understanding [10]. However, relatively little work has focused on applying HMM in sketch recognition. We believe that it is because temporal information of sketch is not as robust as that of the handwritten characters. Henry and Wardhani [3] proposes to use chain-code like features to recognize isolated symbols. Corners here are detected simply by direction change, and it can only deal with line primitives, while our approach can recognize strokes of both lines and arcs. Sezgin and Davis [9] use HMM to do sketch parsing. Their method can group strokes and recognize each cluster into a distinct symbol through learning of users' drawing styles. Similarly, the method works on symbol level. Since the drawing patterns vary between users, it is difficult to extend to other applications and new symbols. Cates and Davis[2] present an early sketch processing technique that uses both Markov random field (MRF) and belief propagation to do fragmentation. Their graphical model based approach can incorporate context. In his work, each stroke is represented separately as an MRF, but our model can be used to represent all kinds of strokes.

Although different people may scribble the same shape in different strokes and different orders, drawing of sketch itself is temporal related. And we believe the drawing context can help to do fragmentation. In this paper, we present an HMM model that can represent all kinds of strokes making up of lines and arcs. Stroke fragmentation is done by finding out the optimal path of the input with respect to the model. The contributions of our work include: 1). We introduce a model that can be used to represent any arbitrary stroke composed of lines and circular arcs; 2). The model is domain-independent. It does not need any priori information about the constitution of the stroke. Therefore, it can be easily adapted to other applications; 3). Segment point detection and primitive shape approximation are achieved simultaneously. Besides, as it uses both local and the global context, it can deal with smooth strokes properly. In section II, we introduce the states and the model definition. Features and probability density functions (PDFs) are presented in section III. After evaluating the effectiveness of the proposed method, we conclude the paper in section VI.

## 2. DEFINITION OF HMM MODEL AND STATES

### 2.1 Hidden State Definition
For most pen-based application areas, we believe lines and circular arcs are enough to describe a shape. Here, arcs include either open or closed circles. In our approach, lines are grouped into 8 clusters according to their directions; for arcs, they can be divided into arcs on clockwise circle and anticlockwise circle, based on their orientations, as shown in Figure 1. We call all these states ($q^0$-$q^{23}$) basic states. They allow to model the basic primitives used to design a sketch. These different states will produce each a different primitive, while some are quite similar, as for instance $q^0$ and $q^{18}$, which will differ only by the curvature value. To use these generative models to carry out the segmentation, we will have to define what are the local observations. They should be linked of course to local directional and curvature features. Some other states ($q^{24}$-$q^{32}$) will be introduced later; they will be useful to model the switching between the basic models.

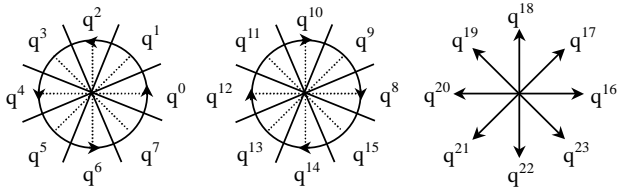

**Figure 1. Definition of primitives and states**

## 2.2 Model Definition and Parameter Selection

In order to construct a model that can represent strokes constituted of arbitrary lines and arcs, we have to concatenate all the states listed above. Figure 2 displays an intuitive solution. With this ergodic model, it is possible to switch from one state to any others. Here all the states are treated equal, not only probabilities of each state to start from, but also probabilities of transition from one state to another. Notice that $q^s$ and $q^f$ are visual states. They can not emit any observation but are used for concatenating purpose. Implementation of this model is easy, and it can represent strokes made up of arbitrary primitives. The problem lies in that it denies the consistency between states of the same primitive. Since when drawing an arc in anticlockwise order, $q^1$ will always follow $q^0$. It implies that the transition cost from $q^0$ to $q^1$ should be smaller than the others. In addition, hand drawings are informal and ambiguous in nature. Model in Figure 2 seems to be kind of noise sensitive.

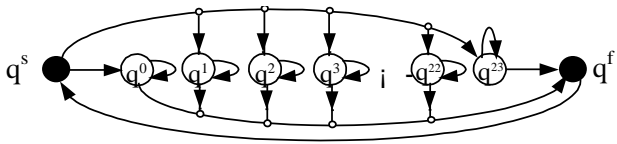

**Figure 2. Generic ergodic model.**

We propose the model shown in Figure 3. It is composed of 3 parts, i.e. model of clockwise circle (top), model of anticlockwise circle (middle) and model of lines in 8 different directions (bottom). Similarly, $q^s$ is non-emitting state, whereas $q^{24}$ and $q^{25}$ can emit all the observations with the same likelihood. State 24 and 25 are used for modelling connecting points between two distinct primitives. These points usually have particular geometry features and cannot be assigned to any of the basic states. Up to two states are allowed, due to the irregularity of sketches and the errors imported in resampling.

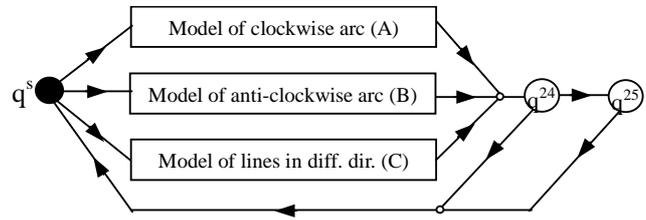

**Figure 3. A sketch map of our model.**

Model of anticlockwise circle (model B) is given in Figure 4. As for model defined in Figure 2, it can start and stop at any state; but the difference is that there are specific transitions connecting adjacent states, which are used to ensure the continuity of the drawing of arc. Model in clockwise order is analogous.

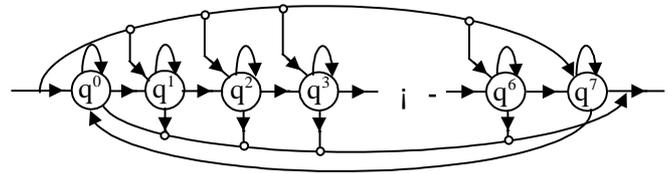

**Figure 4. Model of anticlockwise circle.**

Figure 5 displays the connectivity of the line model (model C). In addition to the 8 basic line states ($q16$- $q23$), there are some special states ($q26$- $q32$) used to model the points on the corner when two distinct lines intersect with each other, for instance $q26$ models the connection between an horizontal line ($q16$) and a 45° line ($q17$), (for simplicity, states between other line states are omitted, actually there is a total of $8 \times 7$ additional states). Curvature measurements for these connection states will be relatively high (computation of curvature can be found in section III), almost equal to that of an arc, although the corresponding point are actually on a line segment. These additional states will be more specific than states 24 and 25 in Figure 3, which can emit all the observations without any preference, whereas these additional states are defined to focus on a specific situation, they will give a higher probability for points on the corner compared with states 24 and 25, so as to provide a better performance.

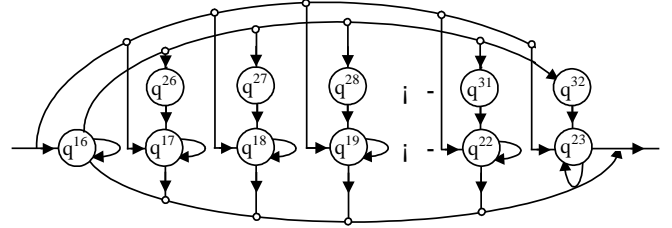

**Figure 5. Model of lines in different orientations.**

In most cases, parameters of the HMM model are obtained through training. Here, all the parameters we used are based on either prior knowledge or strict mathematical deduction. For simplicity, we make the following assumptions: first, probabilities of each basic state ($q^0$-$q^{23}$) being the first state are all the same; second, self-transitions have the same value than transition to the following state (arc states only); third, transition probabilities from each basic state to the corner state- $q^{24}$, are identical.

Apart from the transition probabilities, we have also to define the observation probabilities. From every points $p_i$ of the sketch, we will derive an observation $o_i$ based on a set of 4 features ($f_1, f_2, f_3, f_4$). They are presented in the next section, which will allow the computation of the likelihood of the observation $o_i$ emitted by the HMM. As we assume the independence of features, the global pdf will be defined by :

$$pdf(o_i | q^j) = \prod_{k=1}^{4} pdf(f_k | q^j) \qquad (1)$$

## 3. FEATURE COMPUTATION

### 3.1 Feature Selection

Two main features have been considered, namely local direction and local curvature. They will allow distinguishing the different states of the models. Directions are easy to be computed. As for lines, direction of $p_i$ is that of line $p_{i-1}p_{i+1}$; and for arcs, it can be proved that after resampling direction of the tangent line of $p_i$ is the same as that of line $p_{i-1}p_{i+1}$.

Cosine and sine value of the slant angle of line $p_{i-1}p_{i+1}$ can be described with the coordinates of these two points; while definition and computation of curvature is much more intricate. When dealing with discrete data, curvatures are always computed according to the distribution of points falling into a given area. Here, we denote curvature as the distance of the point to the chord of its neighbouring window. For points on a line, the distance will be small, nearly zero; while for points on an arc, the value will be relatively bigger. Figure 6 illustrates the expression and their physical meanings.. Moreover, curvature is signed. It can be either positive- when drawn in clockwise order; or negative- when in anticlockwise order.

$$f_1 \doteq \cos\varphi = \frac{x_{i+1} - x_{i-1}}{\sqrt{(x_{i+1} - x_{i-1})^2 + (y_{i+1} - y_{i-1})^2}}$$

$$f_2 \doteq \sin\varphi = \frac{y_{i-1} - y_{i+1}}{\sqrt{(x_{i+1} - x_{i-1})^2 + (y_{i+1} - y_{i-1})^2}}$$

$$f_3 \doteq curvature(p_i) = h = distance(p_i, p_{i-2}p_{i+2})$$

$$f_4 \doteq \psi = \pi - angle(p_{i-1}p, pp_{i+1})$$

(a) Expression of feature computation.

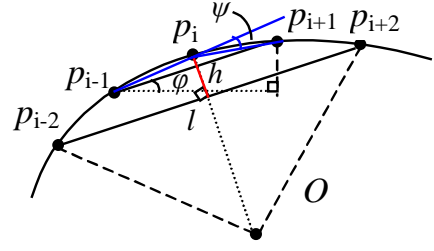

(b) Illustration of the features.

**Figure 6. Expression of feature computation and their physical meanings.**

Size of window should balance between lines and arcs. Ordinarily, the bigger the window is, the more insensitive it will be to the noise. Yet at the same time, it may result in over-smoothing due to the inappropriate use of context; on the other side, a smaller window will be less robust, and less discriminative, but more precise. In our system, we use a window of size 5 points after resampling, where the resampling step is chosen to be adapted to the sketch itself after an analysis of the stroke lengths histogram.

Besides direction and curvature, we import another measurement, namely direction change. It can also be taken as an evaluation of the curvature. As the computation of $f_4$ takes into account of only the two neighboring points, we believe it can help to balance the influence of too much context raised by curvature.

### 3.2 PDF definition

Definitions of pdfs are based on the most widely used Gaussian distribution. Take $f_2$ of $q^{16}$ for example, its slant angle ranges between [-π/8, π/8]. When being in $q^{16}$ the likelihood of $f_2$ is maximum when the angle is zero; as it extends to both sides, the likelihood decreases gradually. The smaller the deviation is, the bigger the pdf. Again, since function |sin(x) – sin(0)| is symmetric in [-π/8, π/8], hence the pdf_$f_2$ of $q^{16}$ is like Figure 7(a): it achieves the maximum value in sin0, and is symmetric to both sides; As for $f_1$ of $q^{17}$, the angle starts from π/8 to 3π/8, with π/4 being the center. Because |cos(x)-cos(π/4)| is asymmetric on the two sides of π/4, so the pdf_$f_1$ of $q^{17}$ is a combination of two gaussian curves (shown in Figure 7(b)). It achieves the peak value in cos(π/4) and decrease to both sides. Moreover, value of cos(π/8) and cos(3π/8) are identical. Pdfs of the other states, and pdfs of direction change can all be defined in the same way.

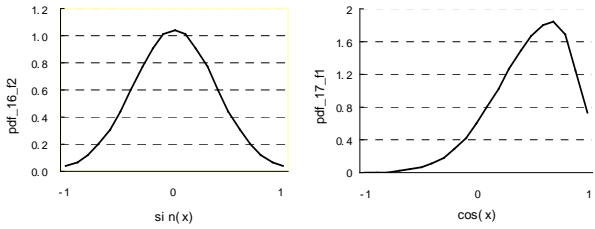

(a) pdf_sine of state16.  (b) pdf_cosine of state

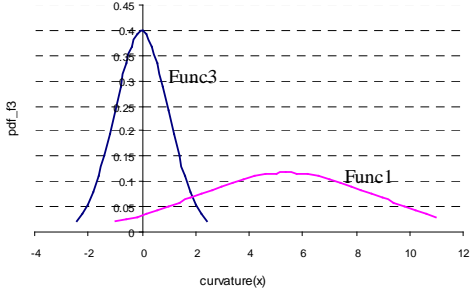

(c) pdf_curvature definition.

**Figure 7. Examples of pdf definition.**

Concerning the pdfs of curvature feature $f_3$, there are three different functions according to which state is concerned. Func1 is for arc states with positive curvature values ($q^8$-$q^{15}$), Func2 is for arc states with negative curvatures ($q^0$-$q^7$), and Func3 is for all the line states ($q^{16}$-$q^{23}$). Since the bigger the radius is, the less the curvature will be. However, if the radius is big enough, it will be difficult to distinguish lines from arcs with only the curvature. Therefore, we constraint the radius of circle range from 10 to 50 resample points respectively, and we believe it is enough for most of the applications. Through experiment, we notice that curvatures of lines remains in the range [-d/8,d/8], and from geometric constraints due to how h is computed, it is bounded by [-2d, 2d], with d being the resampling distance. Figure 7(c) gives the definition of pdf_ $f_3$. Here, we show only func1 and func3. Func2 is symmetric with func1 with respect to the y-axis. From Figure 7(c), the bigger the curvature, the smaller func3 will be. But as when two lines intersect with each other, points on the corner will have a big curvature, which means for line states it should also emit a big curvature. So during computation, when curvature is beyond a threshold, func3 will be assigned with a comparatively bigger value.

## 4. STROKE FRAGMENTATION

Pen-based devices are usually based on isochronal sampling. Computation on the original data will be either time consuming or noisy. Hence, we introduce resampling to separate strokes into equally distributed segments. The resampling distance will be critical for fragmentation. As for lines, the distance is the smaller the better. Since under a tiny distance, a small fluctuation will not influence much for the curvature. But for arcs, points fall into a small area will look like those on lines. Besides, it will be much easy to preserve consistency of an arc if the resampling distance is bigger. At this stage it is important to define a sampling distance which will be related to the scale of the sketch and not to the parameter of the device.

Since for HMM, each pattern should have at least 4 to 5 observations, we finally define the resampling distance to obtain these number of points for the smallest primitives based on an estimation of their length..

The process of HMM-based stroke fragmentation are as follows. First, resampling is introduced to get feature points; then points are transferred into model recognizable observations through feature computation. The Viterbi algorithm is employed to output the optimal state sequence. Take an "L" like stroke for example, based on our model, the output of the segmentation will be something like "22, ..., 22, 68, 16, ..., 16". The last stage of this segmentation method is a post-processing, which aims at fine-tuning the localization of the segmentation points, and an optional color label can be assign to visually assess the results of the segmentation. First, in the previous example, point 68 will be taken as the candidate segment point, and then the point having the maximum curvature nearby is selected to be the final result. We do not take the point corresponding to q68 to be the exact segment point, because we cannot guarantee that after resampling points will cover all the real segment points. But we can be sure that the real segment points are in the neighborhood of the feature points. At last, the different types of primitives are displayed in different color.

## 5. SYSTEM EVALUATION

We selected several examples from the existing stroke fragmentation researches to evaluate the performance of our approach. During selection, we try to avoid those that can have different segmentation results. Besides we added example 9-12, to test the usability of our method in segmenting handwritten digitals and characters. Figure 8 lists all the examples. Small circles indicate the real desired segment points. As the start point and the end point are treated as segment points automatically, they are not included in our evaluation.

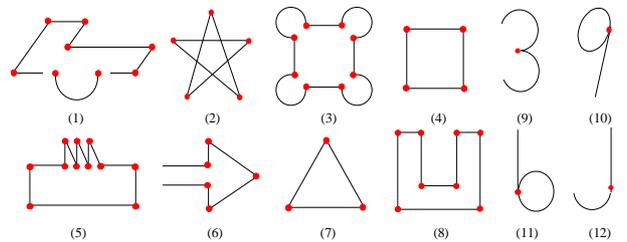

**Figure 8. Samples used in our experiment**

We asked 6 people to draw the examples. Before experiment, we first introduced them the input device, and then ask each of them to sketch freely for about 5-10 minutes to learn how to use it as naturally as a pen. They are told the purpose of our experiment and primitives the system can recognize, in order they will not draw too casually. People are asked to draw each of the example 8 times discontinuously. There is no constraint about the direction and the size. But as some of the drawings are far from the objective, in order to guarantee the veracity of the evaluation, they are unused. At last, we got 588 valid samples, which are all used for testing.

Model shown in Figure 2 is implemented to serve as a baseline approach. Experiment results are given in Figure 9. The two measurements are false-negative rate and false-positive rate. The definitions are given below.

$$false\_positive = \frac{no\#\_accepted\_false\_segment\_point}{no\#\_all\_accepted\_segment\_point} \times 100\%$$

$$false\_negative = \frac{no\#\_rejected\_true\_segment\_point}{no\#\_all\_true\_segment\_point} \times 100\%$$

From the experiment it can be seen that for all the users, our approach has achieved a better performance than the baseline method. The false negative rate in our approach is very low, less than 1%. It means our system can locate almost all the real segment points. While the false positive rates vary between users. We found that most of the false positive occurs when an arc is over segmented, which is also the cause of low recognition rate of arcs. As HMM-based recognition tries to find out the optimal matching with model based on dynamic programming. So if user draws a long and flat segment when sketching an arc, the probability of recognizing it as a line is far more bigger than that of an arc. This argues that users' drawing may deviate from their exact intention. Some other false positive happens if there is a big direction change when drawing a straight line. Further, our method has a poor performance on user2, this is because most arcs he drew have exceeded our predefined scope. False negative arises when dealing with sample 5. Since when user draw the broken lines too small, it will make curvatures of points on the corner be comparatively bigger, and results it to be recognized as arc. We argue that these kinds of mistakes can be avoided if we explain to users the range we can handle before experiment. Figure 10 shows some examples of the correctly segmented strokes and some failed examples (Small round circles are the segment points, and primitives are displayed in different color).

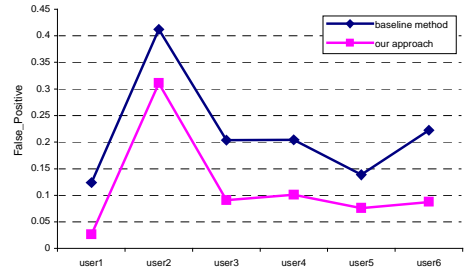

**(a) recognition result of false positive.**

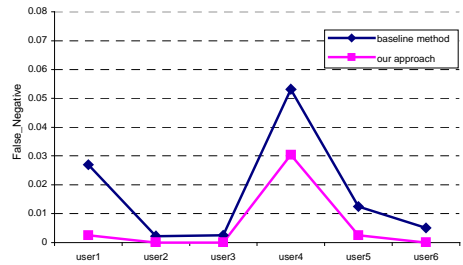

**(b) recognition result of false negative.**

**Figure 9. Recognition results.**

We found that the performance on 1-8 is a little better than on 9-12. This is due to the fact that most users will draw a shape according to the template, but the drawing of digitals and characters will be influenced by more subjective factors. Again, the symbols themselves are not made up of regular arcs, which is another cause of over-segmentation. However from the experiment we realize that due to the management of context, our method can achieve better results for smooth strokes.

As for processing time, HMM-based methods has a computation complexity of $O(N^2T)$. N is the number of states and T is the length of the observations. During our experiment, the average processing time of each stroke is 247.0833 milliseconds. Compared with their drawing interval 2391.917 milliseconds, we believe it can meet the requirements of real-time processing.

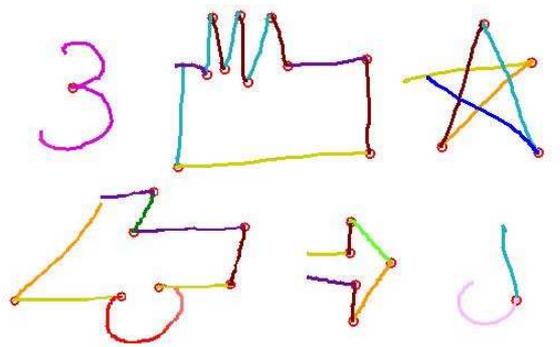

**(a) Successfully fragmented examples**

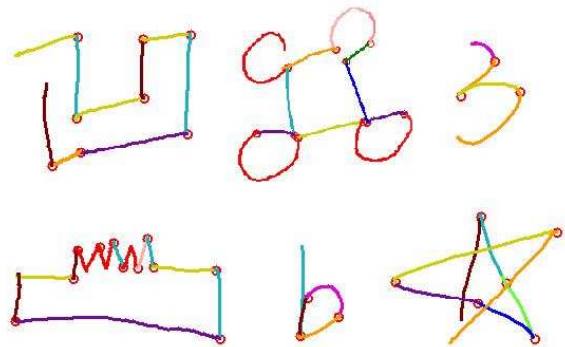

**(b) Fail to segment examples.**

**Figure 10. Recognition examples.**

## 6. CONCLUSION

In this paper, we propose an HMM-based stroke fragmentation paradigm. This is the first attemp at trying to build a model on stroke level. Our model can represent an arbitrary stroke composed of lines and circular arcs. As no priori knowledge is needed, it can be easily adapted to other applications. Besides, our approach can do segment point location and primitive type determination simutaneously. Since it utilize both the local and global contexts, it can efficiently deal with smooth strokes that can not be handled properly by other approaches. The preliminary experiments verify that it cannot only be used to compress data, but also at the same time preserving its geometry characteristics.

We believe context is good to do stroke fragmentation. But on the other hand it sometimes may bring problems. Due to the inconsistency between the continuity in computation of curvature and the discontinuity of curvature between primitives, some short segments in the broken line will be recognized as arcs, which causes the recognition rate of lines to be decreased. Hence, further research can focus on utilizing context information more rationally. Moreover, as sketches are ambiguous in nature, the hand drawn pictures may deviate from users' exact purpose, so some more reasonable measurements are urgently needed to evaluate the performance of a stroke fragmentation system.

**ACKNOWLEDGMENTS**
We thank Pierre-Michel Lallican and Sylvain Togni of VisionObjects Company for their helpful suggestions.